\documentclass{appolb}

\usepackage{graphicx}
\usepackage{xspace}
\usepackage{xcolor}
\usepackage{ulem}
\usepackage{definitions}
\usepackage{amssymb}
\usepackage[mathlines]{lineno} 
\usepackage{cite}

\begin{document}

\title{Missing beauty of proton-proton interactions
\thanks{Presented at the XIV$^{\mathrm{th}}$ International Conference on Beauty, Charm and Hyperon Hadrons, June 5-11, 2022, Kraków, Poland by Alexander Milov}}
\author{Iakov Aizenberg, Alexander Milov
\address{Department of Physics and Astrophysics, Weizmann Institute of Science, Rehovot, 761001, Israel}
\\[3mm]
{Zvi Citron 
\address{Department of Physics, Ben Gurion University of the Negev, Beer Sheva, 8410501, Israel}
}
}
\maketitle
\begin{abstract}
From first principles, particles with the same quark content and similar masses should have similar kinematic distributions. Transverse mass scaling may be employed to estimate possible differences in the momentum distribution of such particles. Based on this scaling the excited bottomonium states measured at the LHC are found to be significantly different from $\Upsilon$(1S) to the extent that the integrated yield of $\Upsilon$(2S) is 1.6 times less and $\Upsilon$(3S) 2.4 times less than would be explained by the mass difference. This proceeding explains how the estimate is worked out and relates it to other measurements performed at the LHC.
\end{abstract}

Recent measurements conducted by the ATLAS~\cite{ATLAS:2022xar}, CMS\cite{CMS:2013jsu,CMS:2020fae} and LHCb~\cite{LHCb:2020sey} experiments show that in proton-proton (\pp) collisions the yield of excited quarkonia states, \equm,  with respect to the ground state, \qum , diminishes with the event multiplicity. The magnitude of the effect increases at low transverse momentum (\pT) of the particle. This effect should have an impact on the \pT distribution of the \equm which can be measured in an independent analysis. An approach to do that can be based on the assumption that particles with the same quark content and close mass should have similar momentum distributions. This assumption is appropriate for the \Upsa-meson family, in which the masses of the \Utwo and \Uthree are larger than the mass of \Uone by 6\% 9.5\%, respectively.

In a broader sense, this assumption is the basis of the transverse mass (\mT) scaling, stating that particles produced in \pp collisions have the same \mT-distribution, where $\mT=\sqrt{\pT^{2}+m^{2}}$ and $m$ is the particle rest mass. The \mT-scaling is widely exploited in many phenomenological studies~\cite{Grigoryan:2017gcg,Bratkovskaya:1997dh,STAR:2006nmo,Altenkamper:2017qot,PHENIX:2010qqf} and is often used in the form 
\begin{equation}
\frac{\mathrm{d}\sigma}{\mathrm{d}\mT} \propto \left(1+\frac{\mT}{nT}\right)^{-n}.
\label{eq:mT}
\end{equation}
The \mT-scaling phenomenon is equivalent to the fact that the exponent $n$ and parameter $T$ are universal for all particles, although different for mesons and baryons (e.g.~ \cite{Altenkamper:2017qot} and references therein). This analysis considers only mesons. The scaling behavior of heavy mesons is studied in comparison to light quark species and is used to understand the ratios between excited and ground quarkonia states measured in experiments.

\begin{figure}[htb]
\centerline{\includegraphics[width=0.7\textwidth]{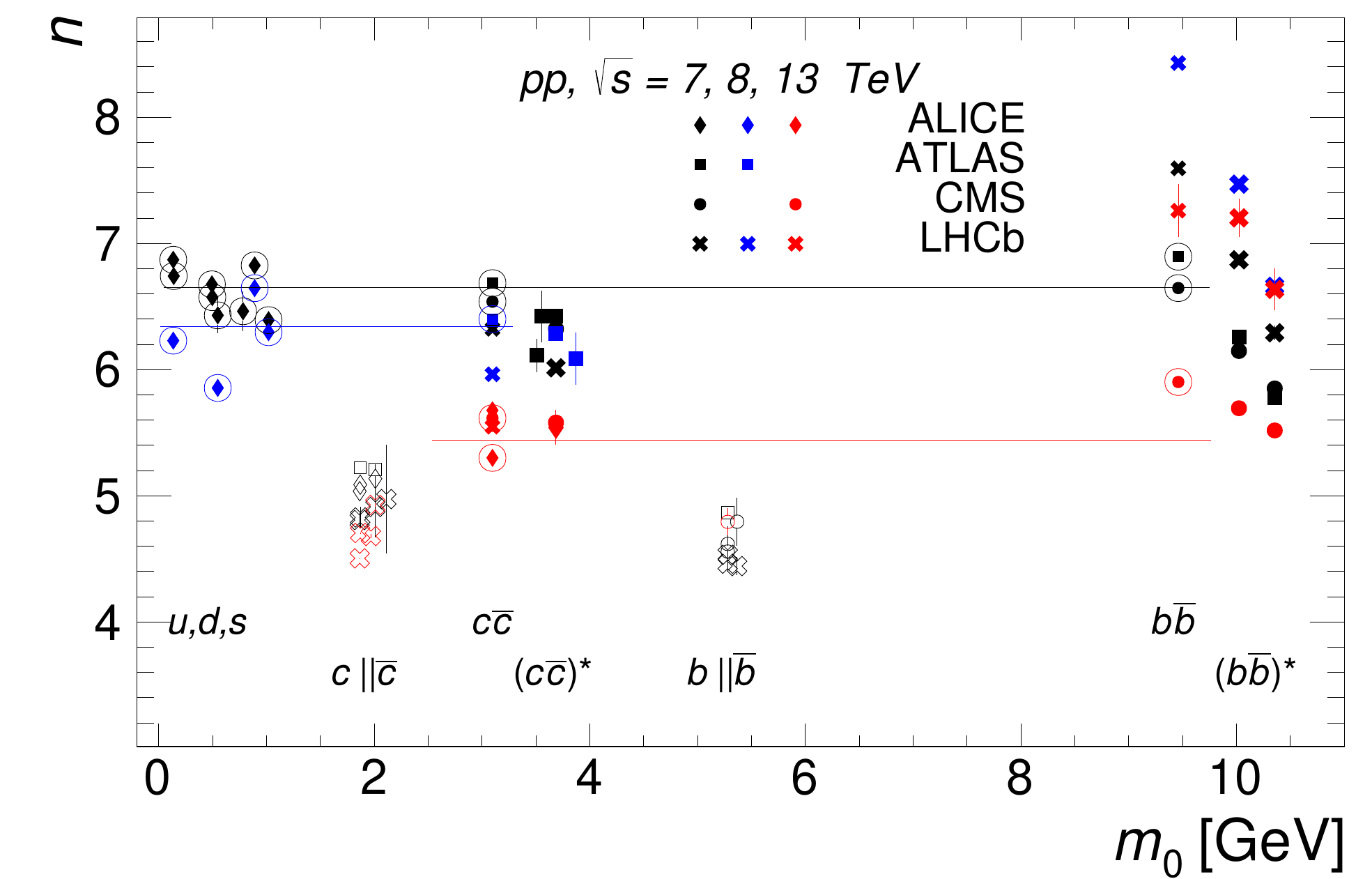}}
\caption{Measured $n$ for different species. Colors correspond to collision energies, symbol shapes denote experiments. Error bars are uncertainties of the fits. Circled points are used for common fits of the \uds and \qum mesons. Results of the fits are shown by horizontal lines. From reference~\cite{Aizenberg:2022iwl}.}
\label{fig:individual_fits}
\end{figure}

A comprehensive study of the \mT distributions of the mesons at three LHC energies $\sqs = 7,8$, and 13~TeV is performed in~\cite{Aizenberg:2022iwl}. It uses 72 data samples of 18 mesons and their isospin partners. Figure~\ref{fig:individual_fits} shows the parameter $n$ obtained in fits of individual data samples. 
When fitting data with the form of Eq.~\ref{eq:mT}, parameters $n$ and $T$ are strongly correlated, therefore $T$ is fixed to 254~MeV. This value is similar to many other studies conducted at LHC energies and it can be somewhat different at different energies. It is shown later that the derived conclusions are indifferent to the exact value of $T$, which is also checked by varying it by $\pm50$~MeV in the analysis.

Several trends are visible in Figure~\ref{fig:individual_fits}. Particles produced in collisions with higher \sqs have harder spectra (lower $n$). The magnitude of $n$ depends on the quark content of the particle. Open heavy flavor mesons (\ohf) demonstrate significantly harder spectra compared to other species and harder for the open bottom (\obtm) than for open charm (\ochm).

At each collision energy, the LHCb results for \Upsa-mesons are above others because they are measured at higher rapidity. Comparing results at the same energy as it is done in~\cite{Grigoryan:2017gcg} shows that $n$ increases with rapidity. Figure~\ref{fig:individual_fits} shows that the exponent $n$ for \uds and \qum mesons are similar. At the same time, \equm have lower values of $n$ than the ground-state:  $n_{\jpsi}>n_{\psits}$, and for bottomonia states: $n_{\Uone}>n_{\Utwo}>n_{\Uthree}$. 

Based on these observations, the values of $n$ for \uds and ground-state \qum mesons measured at midrapidity (circled data points in Figure~\ref{fig:individual_fits}) are fit to extract a common $n$. 
There are 12, 5 and 3 data samples at $\sqs=7$, 8, and 13~TeV respectively. The $\sqs=7$ TeV values are fit to a linear function, which becomes constant for $T=254$~MeV. Due to the low number of selected data sets at higher energies, the 8 and 13 TeV data are initially fit to a constant. The values of $n\left(\sqs=7,8,13\mathrm{[TeV]} \right) = \left(6.65, 6.34, 5.44\right)$, are shown in Figure~\ref{fig:individual_fits} with lines. 

\begin{figure}[htb]
\centerline{\includegraphics[width=0.7\textwidth]{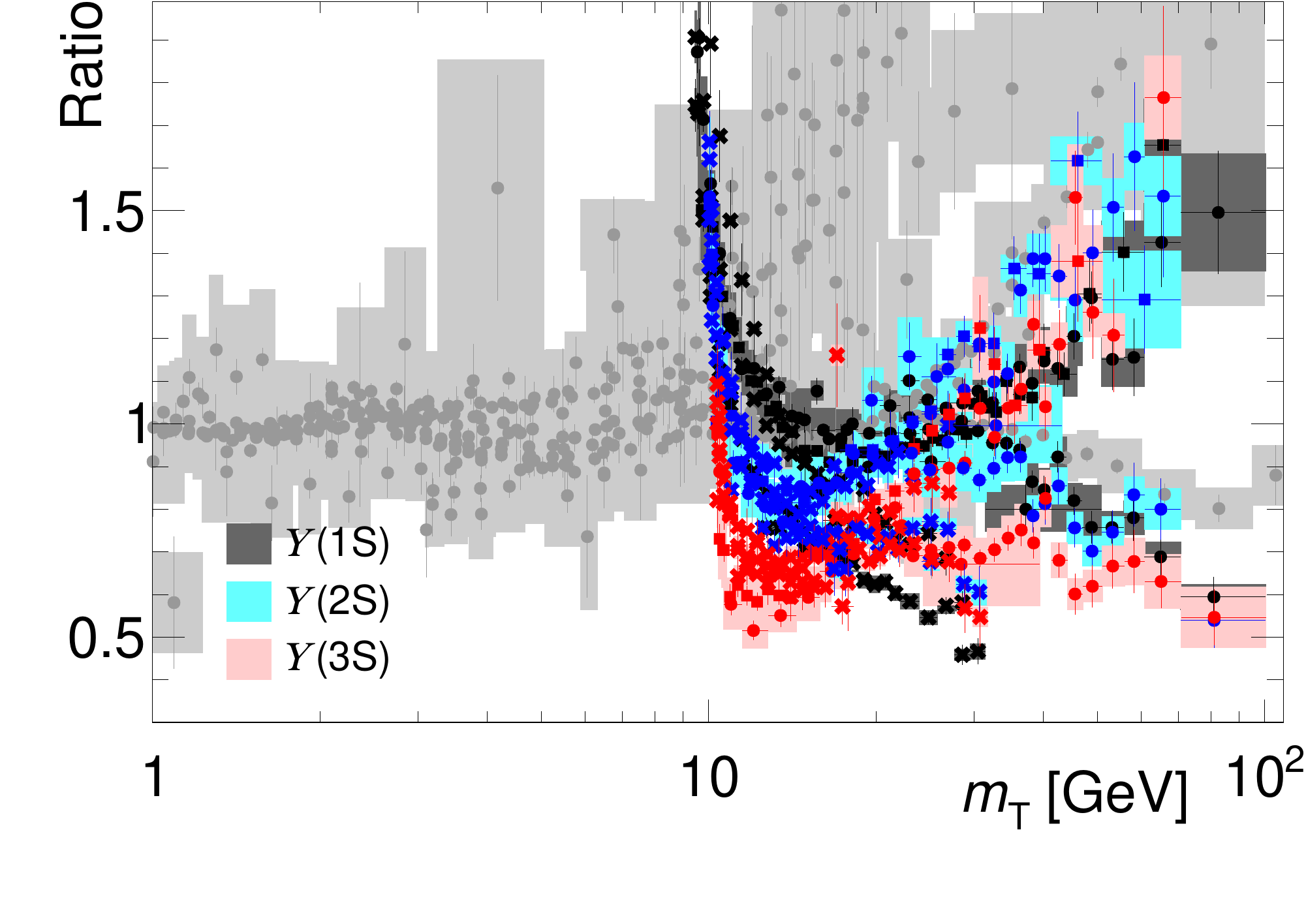}}
\caption{\label{fig:common_fit} Spectra of individual particles divided by the common fit. Measurements used in the common fit are shown in gray. From reference~\cite{Aizenberg:2022iwl}.}
\end{figure}

Figure~\ref{fig:common_fit} shows the \Upsn results at 3 collision 
energies, divided by the common fit. The \Upsa-points are shown in different colors, regardless of energy, and points that are used in the common fit are shown in gray. The latter demonstrate reasonable agreement with unity, although at high-\pT \qum spectra tend to rise up to a factor of 2, somewhat similar to what can be seen in Ref.~\cite{Grigoryan:2017gcg}. Using Eq.~\ref{eq:mT} at high \mT this translates to  
$\mathrm{d}n/n \approx 1/n\times\mathrm{d}\sigma/\sigma\approx15$--18\% deviation of $n$ from the `true scaling' value, which is comparable or even smaller than the difference in $n$ at three measured energies. Similarly, the LHCb results which are measured at high rapidity, yield a value of $n$ that is larger approximately by 1 (see Figure~\ref{fig:individual_fits}) and constitutes approximately the same 20\% deviation from the common $n$. 

\begin{figure}[htb]
\centerline{\includegraphics[width=0.7\textwidth]{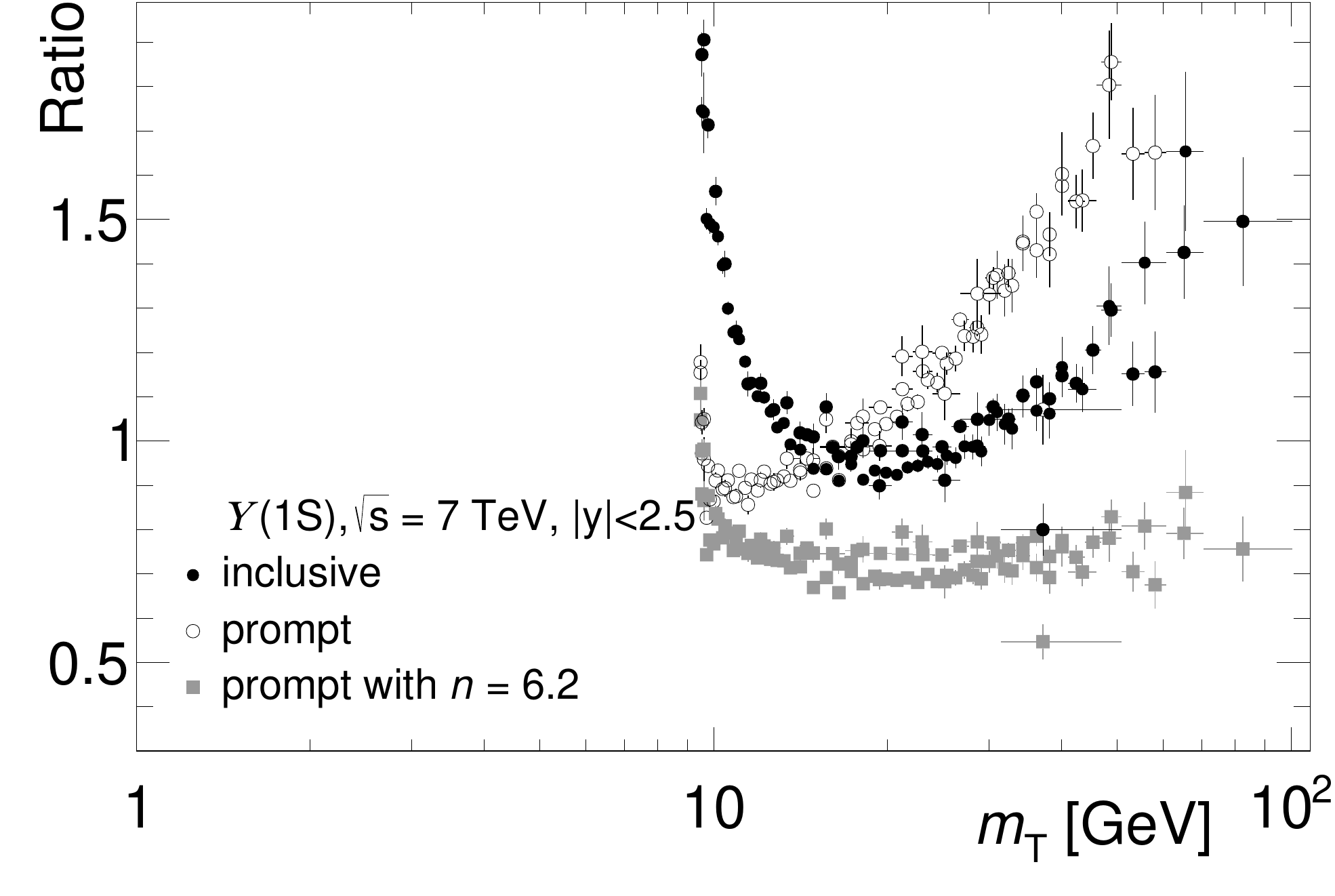}}
\caption{\label{fig:ratio_var} Spectra of \Uone at $\sqs=7$ TeV at midrapidity divided by the common fit. As used in the analysis (full circles), prompt fraction (empty circles), using different $n$ in a common fit (squares). Points are shifted vertically for visibility.}
\end{figure}

A significant rise for all \Upsn states at low \pT, i.e. $\mT\gtrsim m$, is clearly visible. The contribution of $\chi_{b}\mathrm{(1P)}\rightarrow\Uone$ has been estimated using \Pythia \cite{Sjostrand:2014zea} simulations to work out the prompt fraction of \Uone. Inclusive data that is used in the analysis and the prompt fraction, estimated from \Pythia are shown in Figure~\ref{fig:ratio_var} with full and open circles respectively. 
The prompt fraction has only a weak excess at $\mT\gtrsim m$, but it still rises at high \pT. Decreasing $n$ by 8\% flattens the ratio, as shown in the figure with squares. Nevertheless, inclusive data with $\pT>5$~GeV is used in the fit, because prompt fractions for \Upsn are not yet reliably measured experimentally nor fully modeled successfully.

\begin{figure}[htb]
\centerline{\includegraphics[width=0.7\textwidth]{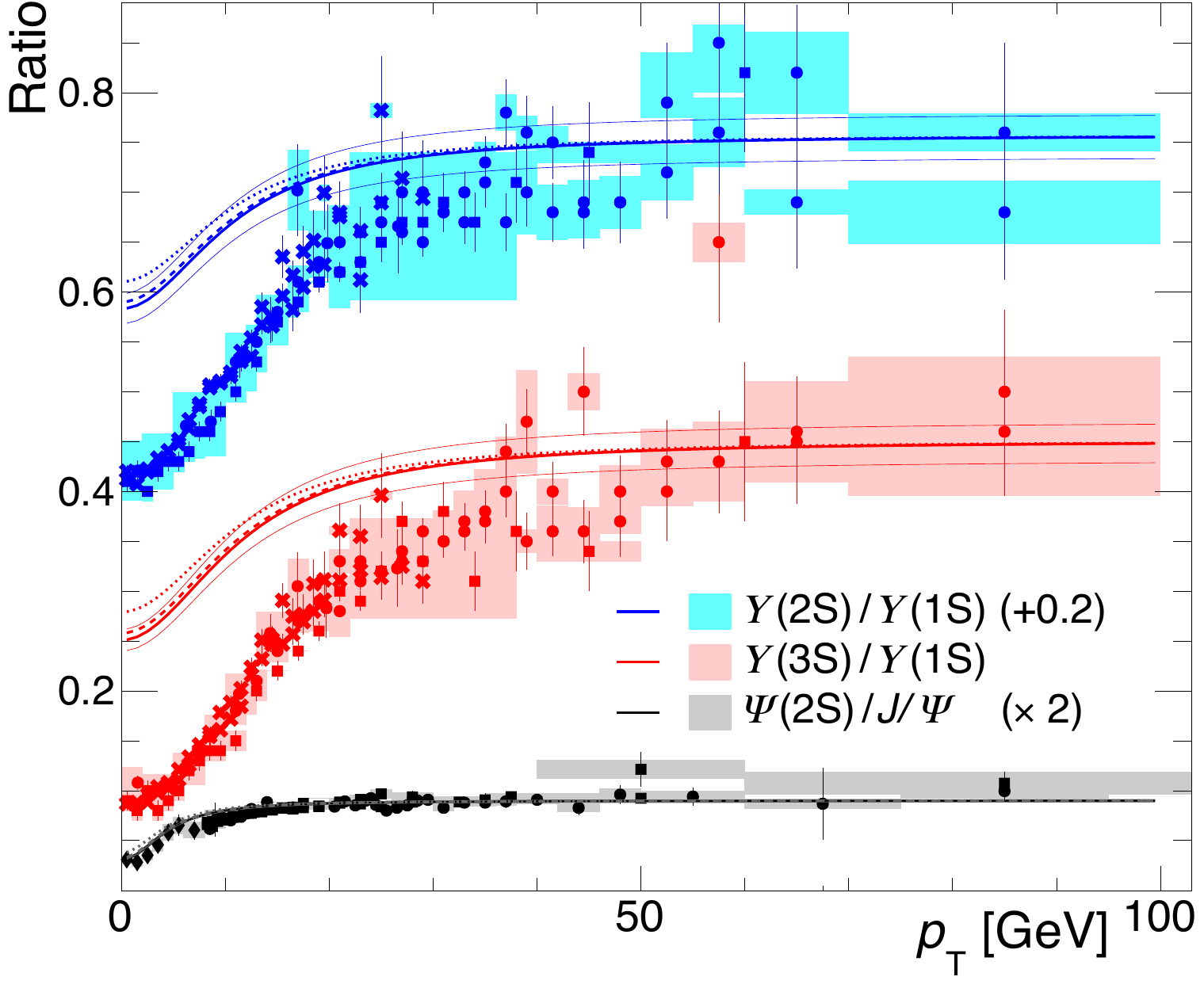}}
\caption{\label{fig:ratios} Measured $\equm/\qum$ ratios (markers) and \mT-scaling prediction normalized to the data at $\pT>50$ GeV with normalization uncertainties (solid lines). The dashed (dotted) lines correspond $\sqs=8$ (13) TeV. From reference~\cite{Aizenberg:2022iwl}.}
\end{figure}

The ratios of all $\equm/\qum$ measured at the LHC are shown in Figure~\ref{fig:ratios}. 
The ratio curves derived from \mT-scaling are drawn normalized to data at $\pT>50$~GeV, which uncertainties are shown with thin solid lines. Dashed lines correspond to higher energies. The shape of the curves can be directly derived from Eq.~\ref{eq:mT} and is governed by their minimum-to-maximum span:
\begin{equation}
\frac{\mathrm{min}}{\mathrm{max}} \approx 1-\frac{\Delta m}{nT+m}n,
\label{eq:rT}
\end{equation}
where $m$ corresponding to \Uone is equal to 9.46~GeV and $\Delta m$ is 0.9 GeV for \Uthree. Equation~\ref{eq:rT} makes it clear that the value of $T$ has only a minor effect on the shape because $nT\ll m$. Since $\Delta m\ll m$, even for \Uthree, the ratio before $n$ in Eq.~\ref{eq:rT} is $\approx0.08$, therefore the two curves for $\sqs=7$ and 13~TeV are close to each other. As discussed above, residual non-flatness of the common fit and the rapidity dependence have similar or smaller impacts on $n$ compared to \sqs. None of these effects can explain the drastic difference between the experimentally measured \ebtum/\btum ratios and the curves shown in Figure~\ref{fig:ratios}. To reconcile $\Utwo/\Uone$-ratio curve with the data, the feed downs from \Pythia that are consistent with LHCb results~\cite{LHCb:2014ngh} and nearly eliminate the peak at $m  \lesssim \mT$ in Figure~\ref{fig:ratio_var}, should be increased by 2.5 and even more than that for \Uthree, which is not plausible.

\begin{figure}[htb]
\centerline{\includegraphics[width=0.57\textwidth]{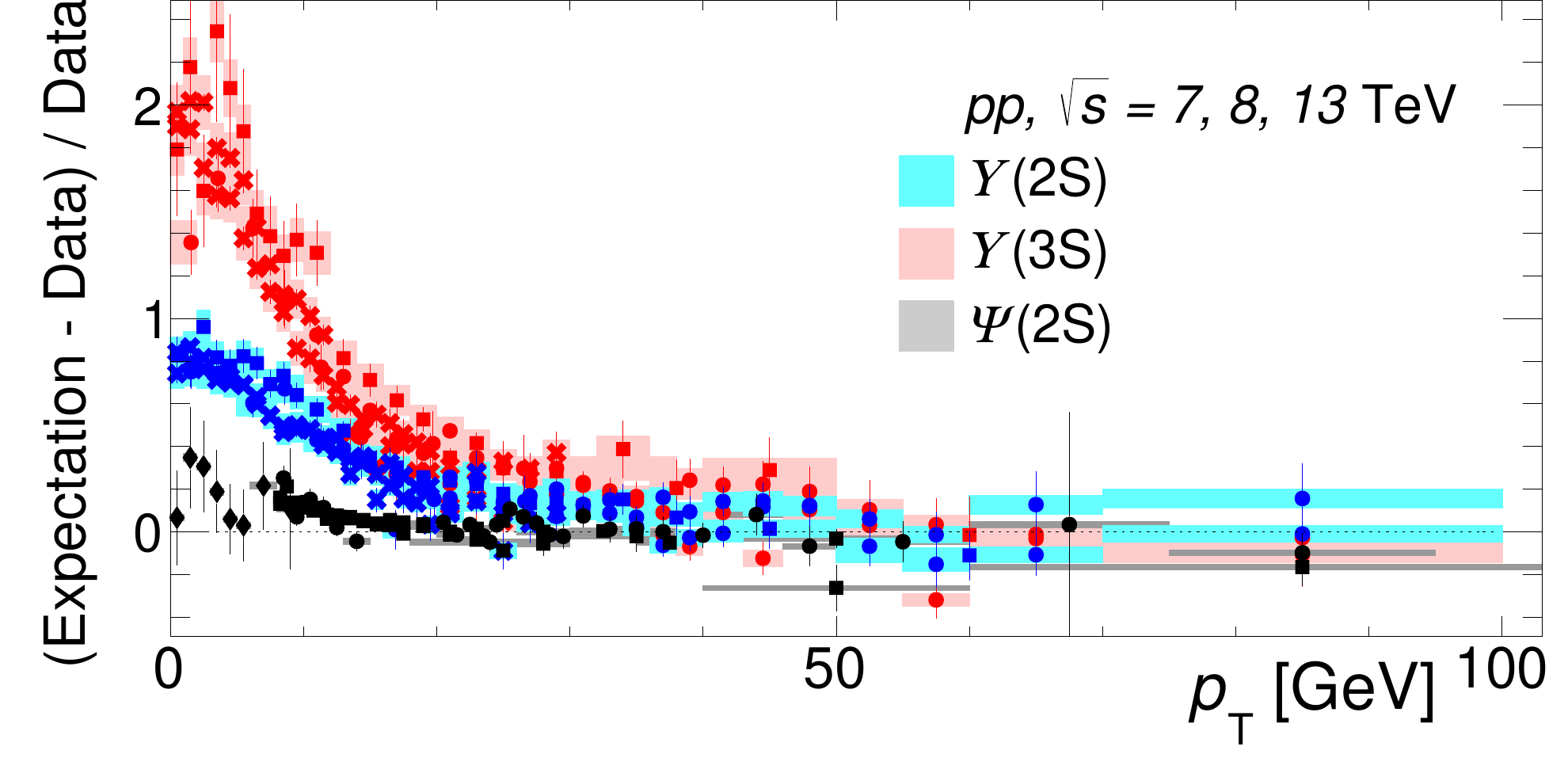}
\includegraphics[width=0.42\textwidth]{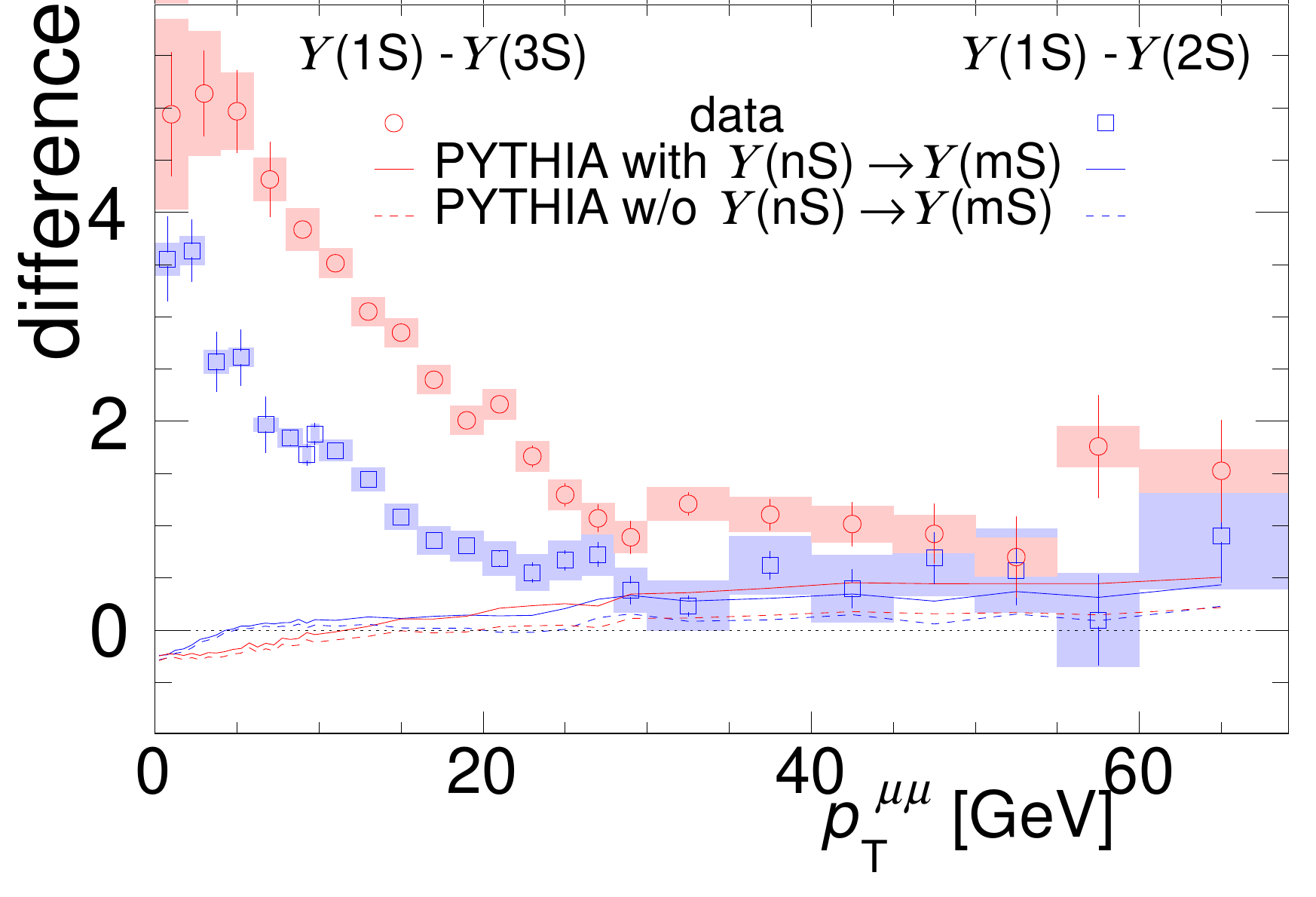}}
\caption{\label{fig:missing} Left: `missing fraction' of \ebtum from reference~\cite{Aizenberg:2022iwl}. Right: difference in the number of tracks in events with \Upsn states from reference~\cite{ATLAS:2022xar}.}
\end{figure}

To quantify the discrepancy between the data and the \mT-scaling prediction one can build the particles' `missing fraction': $\equm_{\mathrm{expected}} / \equm_{\mathrm{measured}}-1$. This fraction is shown in the left panel of Figure~\ref{fig:missing}. Assuming the \mT-scaling scenario and the \Uone-meson production cross-section, expected \Utwo production at low \pT is approximately twice higher than the measurement and \Uthree is roughly three times higher. 

The discrepancy between the measured production rates of \ebtum in \pp collisions and rates expected from the \mT-scaling approach may share a common origin with the results reported in~\cite{Aizenberg:2022iwl,CMS:2020fae}. This would explain the striking similarity between the `missing fraction' and the difference in the number of tracks in the underlying event measured by the ATLAS experiment. This comparison is demonstrated in the two panels of Figure~\ref{fig:missing}. It would mean that the track present in \pp collisions reduces the production rate of \ebtum. Using the `missing fraction' and the measured differential cross section of \Upsn states (for the full list of references see~\cite{Aizenberg:2022iwl}) one can estimate that the measured cross-sections of \Utwo and \Uthree mesons are `suppressed' by factors of 1.6 and 2.4 compared to the expectation from the production rates of \Uone. 

Research of Z.C. is supported by the Israel Science Foundation (grant 1946/18). Research of I.A. and A.M. is supported by Israel Academy of Science and Humanity and the 
MINERVA Stiftung with the funds from the BMBF of the Federal Republic of Germany.

\bibliographystyle{unsrt}
\bibliography{mtproc}
\end{document}